\documentclass{article}

\usepackage{amsfonts}
\usepackage{amsmath}
\usepackage{graphicx}
\usepackage{epsfig}
\usepackage{color}
\usepackage{url}

\title{Senescence, change, and competition: when the desire to pick one model harms our understanding}

\author{Andr\'e C. R. Martins\\
	 GRIFE-EACH, Universidade de S\~ao Paulo\\
	 amartins@usp.br
}

\begin{document}
	
	\maketitle

	\begin{abstract}

		The question of why we age is a fundamental one. It is about who we are, and it also might have critical practical aspects as we try to find ways to age slower. Or to not age at all. Different reasons point at distinct strategies for the research of anti-ageing drugs. While the main reason why biological systems work as they do is evolution, for quite a while, it was believed that aging required another explanation. Aging seems to harm individuals so much that even if it has group benefits, those benefits were unlikely to be enough. That has led many scientists to propose non-evolutionary explanations as to why we age. But those theories seem to fail at explaining all the data on how species age. Here, I will show that the insistence of finding the one idea that explains it all might be at the root of the difficulty of getting a full picture. By exploring an evolutionary model of aging where locality and temporal changes are fundamental aspects of the problem, I will show that environmental change causes the barrier for group advantages to become much weaker. That weakening might help small group advantages to add up to the point they could make an adaptive difference. To answer why we age, we might have to abandon asking which models are correct. The full answer might come from considering how much each hypothesis behind each existing model, evolutionary and non-evolutionary ones, contributes to the real world's solution.

	\end{abstract}

	\section{Introduction}
	
	Tentative, valid theories for senescence are plenty. That aging might provide advantages to the species, even while harming the individual, is something that was proposed already in the XIX$^{th}$ century \cite{weismann1889a}. And that is an idea still defended now \cite{Libertini2019}. On the other hand, when group selection effects oppose individual self-interest, individual advantages tend to cause a much stronger adaptive pressure. While group selection can work under the right circumstances, its weaker characteristic has led many people to look for non-evolutive answers to the ageing puzzle. Proposed explanations include the simple accumulation of unfixed damage over time \cite{medawar52a}. Another possibility is that the genes that cause senescence might have benefits that appear earlier in life \cite{williams57a}. Or it might be the case that our bodies can make better use of the energy needed to repair the damage caused by time at other activities, such as reproduction \cite{kirkwood77a,kirkwoodaustad00}.

	Of course, aging, as everything in biology, must be subject to evolution laws. The actual question is if it is an adaptation in itself. That is, if senescence brings advantages that compensate its flaws, or if it is the other way around. The idea it might provide some advantage has been explored for many possible scenarios. For example, it is clear now that species that age can evolve faster. That is, senescence can promote evolvability \cite{layzer80a,goldsmith03a,mitteldorf04a,bredesen04a,mitteldorf10b,mitteldorfmartins13a,Veenstra2020}. And, according to existing models, ageing might not provide just one advantage, but several. Models show it can help with problems such as elimination of infertile individuals \cite{travis04a,dythamtravis06a}, stabilization of environments and avoidance of extinctions \cite{mitteldorf06a}, and protection against epidemics \cite{mitteldorfpepper09a}.

	There is also strong evidence that sexual reproduction and assortative mating might be important  \cite{Lenart2018}. Indeed, many factors can influence the outcome of evolutionary models of senescence. How individuals disperse over a landscape \cite{Galvan2018}, the existence of non-inheritable traits in a spatial context \cite{Yang2013} as well as possible bandwagon effects that benefit lineages \cite{Solon2019} are a few of the examples.

	Good adaptations must bring benefits. However, determining which adaptations will survive does not depends only on those advantages. It also depends on the details of how living beings interact. And that includes information on the environment, how competition happens, who gets to interact with whom, and so on \cite{Ronce2010}. I explored some of those possibilities in a paper where the species existed on a spatial landscape and could form groups of relatives. And I also introduced in my model the idea that the environment changes. Whatever is a good adaptation now, but not be so good a few generations ahead \cite{martins11a}. In that model, there was a range in the parameters' values where aging was indeed an adaptation. Senescence tended to allow lineages to adapt faster to the change in the environment. That advantage could be large enough to compensate for the loss of elder individuals. A less realistic simplification of that model where mutations could be detrimental more often than helpful also showed that when species evolve over a spatial landscape, aging can indeed promote better evolvability \cite{mitteldorfmartins13a}.
	Similar results for populations that evolve in space and compete for limited resources were also observed in other more recent models \cite{Werfel2015,Werfel2017}. The fact that evolution happens in space and that conditions change seem to have a very important effect on the evolution of aging.

	Of course, none of those proposals solve the question. We do not have an actual solid model that fits the data well. Arguments comparing data versus theories go both ways, in favour of an evolutionary explanation \cite{Libertini2015,Mitteldorf2016} or against it (or more precisely, against programmed aging) \cite{Grey2015,Kowald2016}, depending on the preference of the author. The debate on why we age does look like a case where scientists have started defending their personal choices \cite{martins16a}. Very little has been done to create a consensus between the quarrelling fields \cite{Martins2020inpress}. A few mixed proposals do exist, corresponding to the idea that evolution might not be fully responsible for ageing but still influence the rate we get older \cite{Lenart2017}. The fact remains that we do not have, at the moment, a solid, unproblematic theory \cite{Trindade2013}.
	
	In the next section, I will explore variations of my model, where change helps aging to be an adaptation \cite{martins11a}. An essential and correct criticism of that model was that, for senescence to win, one needs to use considerable rates of temporal change and mutation \cite{Kowald2016}. Actual rates in nature are much smaller. To better understand that problem, I will show what happens when introducing the idea that non-inheritable traits exist \cite{Yang2013}. And I will explore how the barrier between individual and group selection might be weakened by the simple existence of change, even when it is small.

	\section{Methods}
	
	The model was implemented using the NetLogo environment \cite{wilensky99a}. The implementation is available at \url{https://www.comses.net/codebases/3c588d89-2c3c-467a-b220-603caba064ed/releases/1.0.0/}. The competition in the model happens over a two-dimensional grid with periodic boundaries. The grid divides the space in square patches. Agents are created all equal, except that there are two types: those who do age and die as soon as they reach age $o$ and those who do not age and only die from other causes. At each time step, all agents reproduce. Their offspring are born at a distance $b$ of their parent. If $b$ is small enough, most newborns will occupy the same patch as their parents. For larger $b$ (but not too large), they tend to start at a neighbouring patch.  Whenever two agents are in the same patch, they compete, and only one survives.
	
	For the competition, each agent has a fitness of $f_i$. When several agents exist on the same patch, the surviving one is picked with a probability proportional to their $f_i$. That is the chance for agent $i$ to survive will be $\frac{f_i}{\sum_{j \in patch}f_j}$. In the original model, \cite{martins11a}, when an agent was born, its fitness was the same as the fitness of its parent plus a possible mutation. The mutation size is given by $m$, and there were equal chances the change would be $-m$, 0, or $+m$. To represent the change in the environment, all agents had their fitness decreased by $d$ at each time step.
	
	Here, the fitness of agent $i$, $f_i$, will be composed of two parts, an inheritable, genetic term, $f_{gi}$, and a non-inheritable part, $f_{ni}$. The effective fitness, used to determine who survives, is the sum of the two terms, $f_i= f_{gi} + f_{ni}$. When an agent reproduces the dynamics described in the previous paragraph is applied only to $f_{gi}$. The non-inheritable terms starts at zero, $f_{ni}=0$.  While the genetic term is fixed and constant for each individual, $f_{ni}$ increases with time.  At each time step,  $l$ is added at $f_{ni}$, for as long as the individual lives. That increase represents the fact that animals get bigger and learn survival strategies. Aside from senescence effects, older individuals should become more effective at surviving.
	
	\section{Results}
	
\begin{figure}
		\centering
		\begin{tabular}{cc}
			\includegraphics[width=0.48\linewidth]{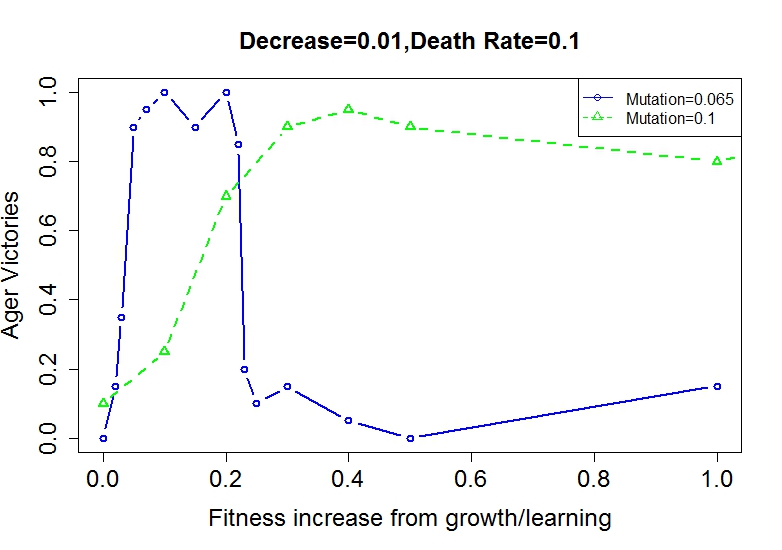}&
			\includegraphics[width=0.48\linewidth]{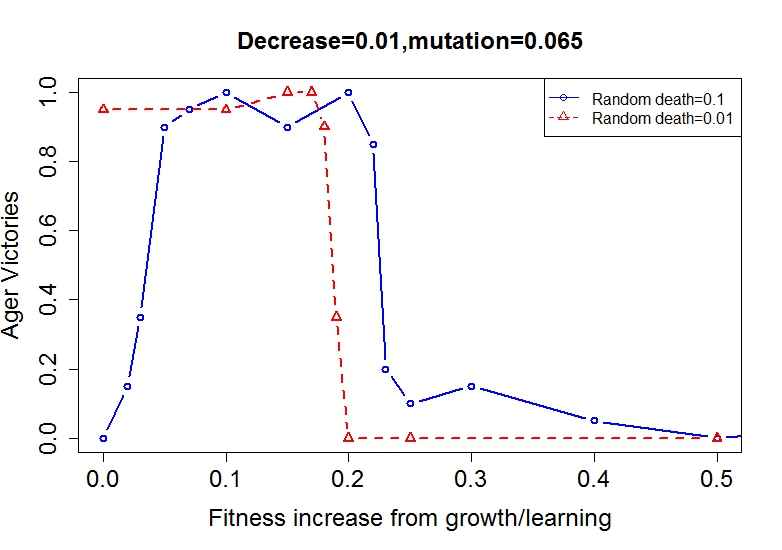}\\
			\includegraphics[width=0.48\linewidth]{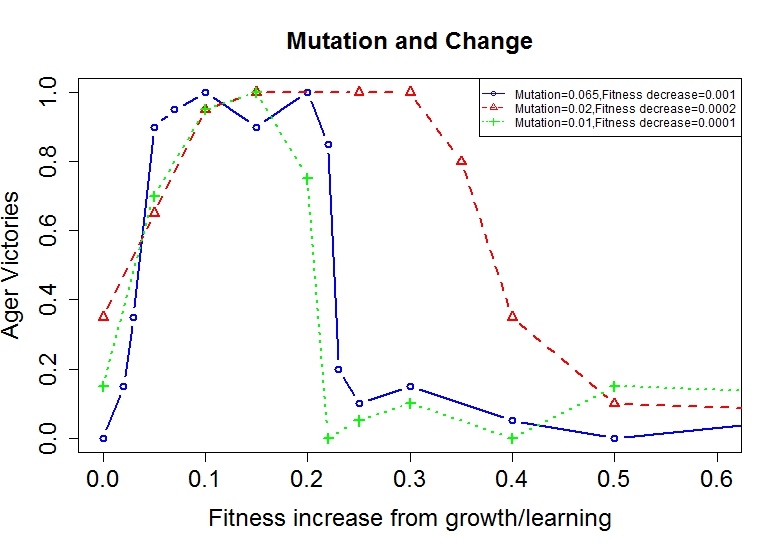}
		\end{tabular}
		\caption{Effects of non-inheritable traits that improve with time (learning, size, etc.) on the proportion of ager victories{\it Top Left}: Case where $d=0.01$ and random death rate of 10\% for two mutation levels. {\it Top Right}: $d=0.01$ and $m=0.065$, for two random death rates. {\it Bottom}: Death rate of 10\%, curves for specific choices of $m$ and $d$
		}\label{fig:noinherit}
	\end{figure}

	In the experiments I ran here, the grid was chosen to go from -30 to +30 in both axes ($61\times 61$).  For all simulations, newly born tended to appear near their parent, as $b=0.5$. For initial conditions, a small cluster of non-ageing agents starts in the first quadrant of the grid, and an equal number of aging agents start at the third quadrant. That allows both sides to form communities before they have to compete with each other. In all simulations presented here, agers die at the age of $o=4$. Figure \ref{fig:noinherit} shows the proportion of agers victories after 20 realizations. It is easy to see how non-inheritable effects often help with the survival of the ager gene.

	In most cases, when learning $l=0$, non-agers win. The exception is when death rate is 1\%, $d=0.01$ and $m=0.065$. However, as $l$ increases, it is easy that the situation changes. Agers start to win. That tendency, however, does not go on for all values of $l$. Instead, we have regions where agers win and, as $l$ becomes even larger, non-agers win again. What happens is that non-agers can accumulate much extra, non-inheritable fitness from $l$. As they grow old, newborns soon have very little chance to dislocate the elders, and non-agers start changing very slowly. On the other hand, agers can not accumulate that much fitness from age, as they die soon. That drives them to adapt much faster. As $l$ becomes too large, the oldest non-agers can become too strong too fast and, while genetically ill-adapted, they can compensate with a strong, effective fitness.
	
	\begin{figure}
		\centering
		\begin{tabular}{cc}
			\includegraphics[width=0.4\linewidth]{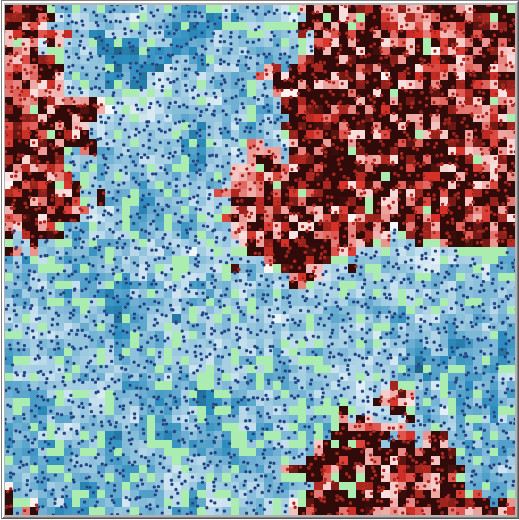}&
			\includegraphics[width=0.4\linewidth]{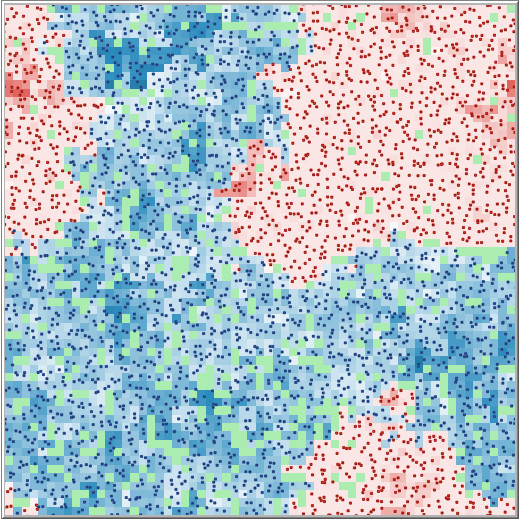}\\
			\includegraphics[width=0.4\linewidth]{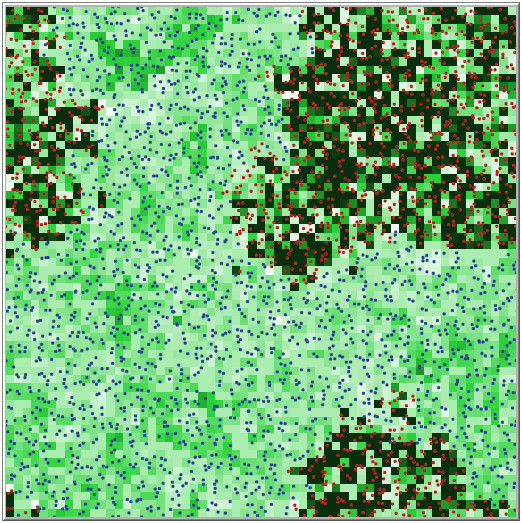}&
			\includegraphics[width=0.4\linewidth]{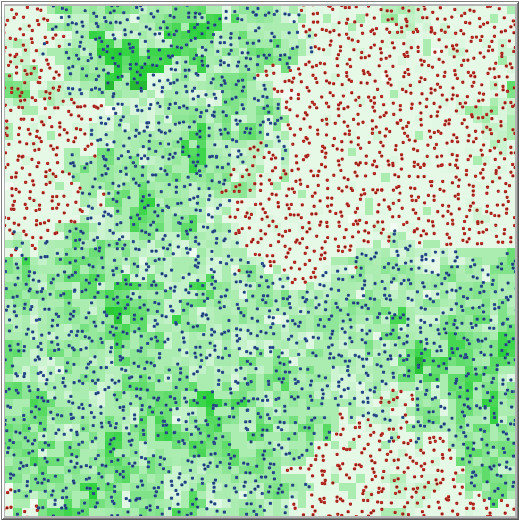}
		\end{tabular}
		\caption{Landscape of competitors in the middle of a simulation where $l=0.1$. The same instant is shown in all figures, darker shades correspond to a larger fitness. The top line shows agents as blue for agers and red for non-agers, while they are all green at the bottom line, for comparison. In the first column of both lines, we see the effective fitness of the agents. At the second column, the genetic fitness.
		}\label{fig:landscapes}
	\end{figure}

	That situation can be easily visualized in Figure \ref{fig:landscapes}, where we can observe a picture of the landscape during the competition between agers (in blue at the first line) and non-agers (red). The case show there corresponds to $l=0.1$, $d=0.01$, $m=0.065$, and a death rate of 2\%. The first column in both lines shows the relative strength of effective fitness. Darker shades correspond to larger values of fitness. The second line brings all agents in green for ease of comparison. It is evident that non-agers have a much larger effective fitness, caused by the constant increase brought by non-inheritable traits over a long life. The second column shows the same instant but with genetic fitness information. In that case, one can easily see that, genetically, the agers are much better. Non-agers win in direct competition, but their offspring are born with a much smaller fitness.
	
	The existence of non-inheritable traits seems to help the survival of agers up to a point. However, the cases in Figure  \ref{fig:noinherit} still correspond to swift changes in the environment and substantial mutation rates. Indeed, massive environmental changes do require large mutation rates; otherwise, every fitness goes to zero. The implementation here includes a minimum very small fitness limit to prevent exactly zero or smaller values. Nevertheless, the dynamics break in a non-realistic way as everyone tends to that minimal value. It is easy to see a minimum amount of mutation $m$ is required for each decrease rate of $d$. 
	
	That is also true as we explore smaller values of $d$. Those values do allow for smaller, more realistic $m$. That means it makes sense to decrease both values together. To understand what happens when changes do become minimal, a series of cases was prepared for small values of $d$. More specifically, $d$ was chosen to be zero (for comparison) as well as $d=1.0 \times 10^{-4}$ and  $d=2.0 \times 10^{-4}$. Mutation rates were chosen for each case so that the effective fitness of agers did not go to zero nor exploded, and each case was repeated for different values of learning rates $l$. Non-agers won in almost every run scenario, except for random death rates of 2\%, $d=2.0 \times 10^{-4}$, $m=0.02$, and only when $l>1.0$.
	
\begin{figure}
		\centering
		\includegraphics[width=0.68\linewidth]{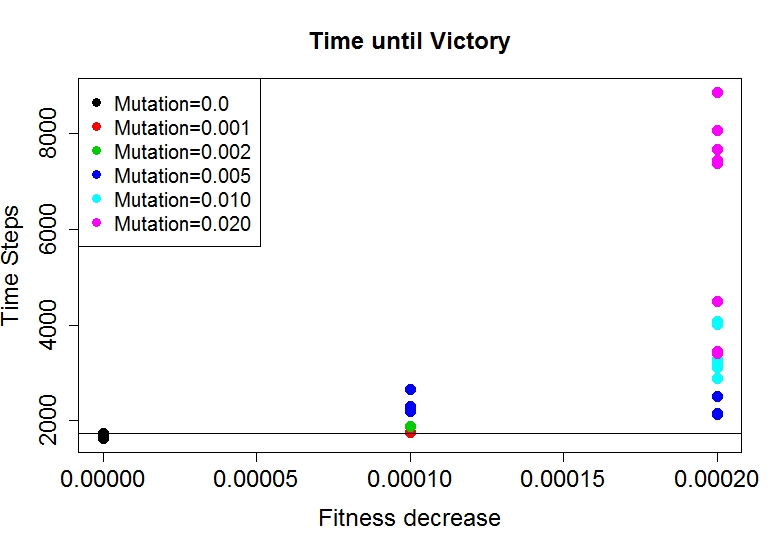}
		\caption{Average time until one of the sides get extinct, averaged over 20 realizations. 
		}\label{fig:times}
	\end{figure}

	Agers can not win for such a small, more realistic amount of environmental change. However, something interesting also was observed. The mere existence of change made the advantage of non-agers smaller. That becomes obvious when the time non-agers take to drive agers to extinction is observed, as in Figure \ref{fig:times}. In the figure, each circle corresponds to the observed time until one of the sides got extinct, averaged over 20 realizations, for a death rate of 2\%. The horizontal line at the bottom corresponds to the case where $d=m=l=0$, shown for comparison. Multiple cases with the same color correspond to distinct values of $l$.

	As soon as we have some temporal change (small $d$), it takes a longer time for a winner to appear. That tendency becomes quite strong when  $d=2.0 \times 10^{-4}$ and increasing $l$. Even when the system is in a state where non-agers still win, changes over time make their evolutionary advantage smaller.  
	
	Senescence was supposed to be unable to be chosen by evolution because individual selection pressure is, under most circumstances, much more potent than group pressure. We know that is not true for systems where change happens fast enough. Time change and viscous environments can provide essential advantages that agers groups might explore. However, the decrease in fitness that was needed to allow agers to win was too large.
	
	\section{Discussion}
	
	Here, we have seen that, while non-agers still win when environmental change is slow and closer to what we see in nature, change can still play an important role. It makes the barrier for the ager victory weaker. When we explored what happens when non-inheritable traits are introduced, new parameter ranges over which senescence would win appeared. Even when that did not happen, the time it takes for the non-ager to win could get significantly larger. Furthermore, that happens even for very small amounts of environmental change.
	
	That suggests that we have been looking at the problem of explaining why we age the wrong way. The many models that show senescence can bring several distinct group advantages might not provide a complete answer alone. However, they do suggest that ageing can bring advantages to groups. As change makes the competitive advantage non-agers have weaker, group effects might play a vital role. It might be the case that, if there are several advantages, each of them might not be enough to make a difference. However, together, they might be more than enough to break the weakened barrier of individual advantage.
	
	In particular, it is time to abandon traditional but wrong ways of choosing models. Our desire to choose one winning model can be quite harmful, both from a cognitive and a logical point of view \cite{martins16a}. Instead, it makes sense to consider, when faced with several possible answers, if the actual cause is not multiple. It is reasonable to assume that many of our current proposals might play a role in explaining what happens. Indeed, exploring all possible theoretical combinations is what we should always do \cite{Martins2020inpress}. In the case of senescence, that means also considering that, while evolutionary and non-evolutionary explanations seem to be initially at odds with each, that is not necessarily the case. In principle, senescence might be adaptive by itself, but the way it was implemented by evolution could be by using genes that cause us to grow older and give us early life benefits. Mixed explanations \cite{Lenart2017} should become more common.  Combinations of the models are not only possible. They must always be considered. That is particularly true now that we have seen that the barrier for group selection, in this specific problem, might be much thinner than we expected.

	\section{Acknowledgments}
	 This work was supported by the Funda\c{c}\~ao de Amparo a Pesquisa do Estado de S\~ao Paulo (FAPESP) under grant  2019/26987-2.

	\bibliographystyle{unsrt}
	\bibliography{biblio}

\end{document}